# Millimeter-deep micron-resolution vibrational imaging by shortwave infrared photothermal microscopy

Hongli Ni,[1,#] Yuhao Yuan,[1,#] Mingsheng Li,[1] Yifan Zhu,[2] Xiaowei Ge,[1] Chinmayee Prabhu Dessai,[3] Le Wang,[1] and Ji-Xin Cheng[1,2,3,4,*]

**Affiliations:**

[1]Department of Electrical and Computer Engineering, Boston University, 8 St. Mary's St., Boston, MA, 02215, USA

[2]Department of Chemistry, Boston University, 590 Commonwealth Avenue, Boston, MA 02215, USA

[3]Department of Biomedical Engineering, Boston University, 44 Cummington Mall, MA 02215, USA

[4]Photonics Center, Boston University, 8 St. Mary's St., Boston, MA, 02215, USA

[#]Equal contribution.

*Corresponding author: jxcheng@bu.edu

**Abstract:** Deep-tissue chemical imaging plays a vital role in biological and medical applications. Here, we present a shortwave infrared photothermal (SWIP) microscope for millimeter-deep vibrational imaging with sub-micron lateral resolution and nanoparticle detection sensitivity. By pumping the overtone transition of carbon-hydrogen bonds and probing the subsequent photothermal lens with shortwave infrared light, SWIP can obtain chemical contrast from microparticles located millimeter-deep in a highly scattering phantom. By fast digitization on the optically probed signal, the amplitude of photothermal signal is shown to be 63 times larger than that of photoacoustic signal, thus enabling highly sensitive detection of nanoscale objects. SWIP can resolve the intracellular lipids across an intact tumor spheroid and the layered structure in millimeter-thick liver, skin, brain, and breast tissues. Together, SWIP microscopy fills a gap in vibrational imaging with sub-cellular resolution and millimeter-level penetration, which heralds broad potential for life science and clinical applications.

**One-Sentence Summary:** Shortwave infrared photothermal microscopy allows for millimeter-deep vibrational imaging with sub-micron lateral resolution.

# INTRODUCTION

Probing cellular activities and functions in intact tissue environment is crucial for a variety of biomedical applications such as cancer pathology and drug discovery (*1*). Vibrational microscopy is a powerful tool for studying cellular functions by providing chemical contrast from nutrients, metabolites and other biomolecules (*2*). However, the imaging depth of current vibrational microscopy is not sufficient to map the chemical content in an intact organoid or a tissue without altering the natural microenvironment. Specifically, infrared spectroscopy-based approaches suffer from a strong water absorption which restricts the penetration depth to tens of micrometers (*3*). Spontaneous or coherent Raman microscopy based on visible or near infrared excitation circumvents the water absorption issue. Nevertheless, the large tissue scattering limits the confocal and coherent Raman imaging depth to around 100 μm (*4, 5*). With spatially offset detection of diffusively back-scattered photons, spatially offset Raman spectroscopy (*6, 7*) and spontaneous Raman tomography (*8-10*) can acquire signals beyond millimeter deep in a tissue. Nevertheless, spatially offset Raman and Raman tomography only offer millimeter-level spatial resolution, not sufficient to monitor cellular level activity. In addition, their sensitivity is limited by the extremely small spontaneous Raman scattering cross-section.

The shortwave infrared (SWIR) region (from 1000 to 2000 nm) (*11*) opens a new window for deep tissue imaging with much reduced scattering compared to the visible region and much reduced water absorption compared to the mid-infrared region (**Fig. 1a**) (*12, 13*). Importantly, the overtone transitions of carbon-hydrogen (C-H) stretching vibration (**Fig. 1b**) reside in this window (*14*), allowing deep vibrational imaging. Overtones are high-order harmonics of the fundamental modes of molecular vibrations whose frequencies can be calculated with the equation: $\bar{v} = \bar{v}_1 n - \chi \bar{v}_1 (n + n^2)$. $\bar{v}_1$ represents the fundamental vibration frequency, $n$ is the order of the overtone, and $\chi$ describes the anharmonicity of the vibration mode. For instance, for the $CH_2$ vibration, the fundamental transition corresponds to the absorption at around 3.5 μm, which leads to the first overtone at around 1.7 μm and the second overtone at around 1.2 μm.

Among the various SWIR modalities, diffuse optical tomography can acquire images beyond millimeter deep in a tissue, yet at millimeter-level spatial resolution (*15*). Photoacoustic (PA) imaging achieves a higher spatial resolution by transducer detection of absorption-induced acoustic waves (*16, 17*). Photoacoustic microscopy (PAM) in the SWIR window allowed vibrational mapping of lipids in arterial tissues and drosophila embryo (*13*). Nevertheless, its sensitivity is not sufficient for visualizing small features inside a cell. In SWIR-PAM, the transducer is placed at a considerable distance away from the absorption site. A dramatic signal loss takes place during the propagation as the acoustic energy diffuses in a 3-D manner, which eventually degrades the detection sensitivity and constrain the ability to detect targets smaller than tens of micrometers in the SWIR region. Additionally, in order to minimize the acoustic signal loss, an acoustic coupling medium is required to use between the sample and transducer. This requirement complicates the optical path design and is not applicable to a sample sensitive to mechanical contact such as a patient wound (*18*). Optically probed photoacoustic spectroscopy is developed for remote sensing purpose (*18, 19*) and has been extended to the SWIR window (*20, 21*). However, the sensitivity of photoacoustic remote sensing is not sufficient for subcellular chemical imaging.

Here, we present a shortwave infrared photothermal (SWIP) microscope that offers subcellular resolution, nanoparticle detection sensitivity, and millimeter-deep tissue imaging capability. By optically detecting vibrational overtone absorption-induced temperature rise which leads to a subsequent change of refractive index at a sample (**Fig. 1c**), SWIP prevents the signal loss during propagation and eliminates the necessity of sample contact. The tightly focused SWIR laser used in SWIP provides subcellular spatial resolution and meanwhile allows millimeter-level imaging depth in highly scattering phantoms and tissues. With high-speed digitization of signals, we capture both photothermal (PT) and PA contributions. Strikingly, the optically probed PT signal is found to be 63 times greater in amplitude than the optically probed PA signal. The strong signal level allows SWIP imaging of single 1-µm polystyrene (PS) beads through 800-µm thick scattering medium. Our photothermal dynamics detection scheme further enables the extraction of signals from small objects over background from the surrounding medium. With these advances, we demonstrate volumetric SWIP imaging of intracellular lipids across an intact tumor spheroid, lipids in thick animal tissue slices, and multilayer of fat cells in human breast biopsy.

## RESULTS

### A shortwave infrared photothermal (SWIP) microscope

To fulfill the SWIP concept, we built a microscope whose schematic is shown in **Fig. 1d**. A 1725 nm laser with 10 ns pulse duration and 2 kHz repetition rate serves as the excitation laser, while a 1310 nm continuous wave (CW) laser is used as the probe. The two beams are combined and focused into the sample through an objective. The signal-carrying probe beam is detected by a biased InGaAs photodiode. The photocurrent from the photodiode is converted to a voltage signal with a 50 Ohm impedance, amplified by an AC coupled low-noise voltage amplifier, and then digitized by a high-speed data acquisition card.

The 1725 nm excitation laser is chosen to excite the first overtone of C-H stretching vibrations. Although the first C-H overtone absorption cross section is around two orders of magnitude smaller than the fundamental absorption (*22, 23*), detecting at the first overtone region can circumvent the strong water absorption in the mid-infrared region where water absorption is more than 3 order stronger than that in the SWIR region (*24*). Moreover, compared to the second overtone, the first overtone of C-H gives 7 times larger signal from lipids (*25*).

When overtone absorption occurs, the refractive index (RI) of the absorber is modulated because of the local temperature change. The modulated RI forms a micro-lens and consequently alter the propagation of the 1310 nm probe laser, which is eventually turned into a light intensity modulation by collecting light through a small aperture inside the condenser (**Fig. 1d**). The formed micro-lens should have an appropriate axial offset in relation to the probe focus to maximize the intensity modulation (*26*). The axial offset of the two laser focuses is optimized with a telescope on the excitation beam.

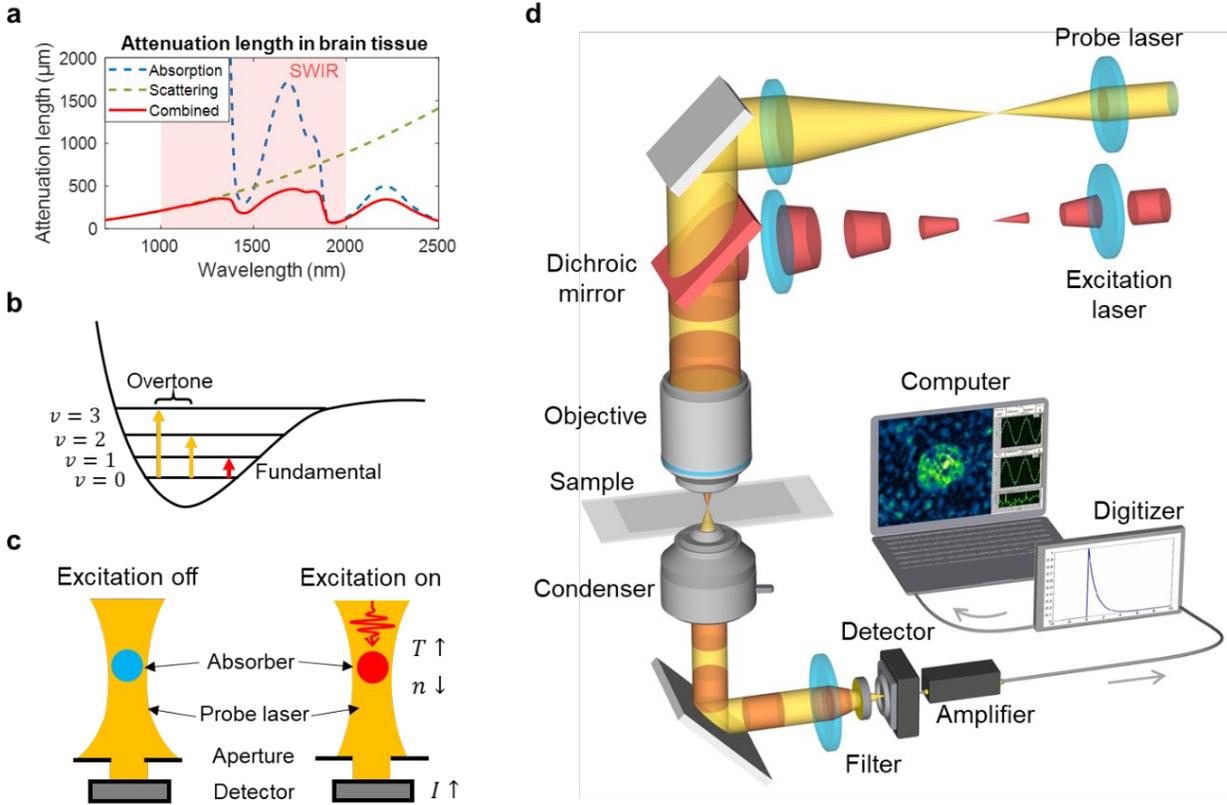

**Figure 1. SWIP microscope principle and schematic.** (a) Wavelength-dependent attenuation length in brain tissue calculated with water absorption and brain tissue scattering coefficients (*12*). (b) Overtone absorption energy diagram. (c) Photothermal contrast mechanism. T: temperature, n: refractive index, I: light intensity. (d) Short-wave infrared photothermal microscope schematic.

**Theory of photothermal and photoacoustic contribution to optically probed signals**

With a pulsed excitation laser, PT and PA conversions occur simultaneously, and both can modulate the RI through the thermo-optic or elasto-optic effects, respectively. In the meantime, optical probing offers an unprecedented detection bandwidth to monitor PT and PA processes in the same microscope. As PT and PA have different signal profiles and characteristics, it is worthwhile to study their relative contributions to the optically probed signal.

The amplitude of PT induced RI change $\delta n_{PT}$ and PA-induced RI change $\delta n_{PA}$ can be written in terms of initial temperature rise $\Delta T$ and initial pressure rise $p_0$ (*18, 27*):

$$\delta n_{PT} = \alpha \Delta T \qquad (1)$$

$$\delta n_{PA} = \frac{\eta n_0^3 p_0}{2\rho v_a^2} \qquad (2)$$

Where $\alpha$ is thermo-optic coefficient, $\eta$ is elasto-optic coefficient, $n_0$ is the initial RI of sample, $\rho$ is density, $v_a$ is speed of sound. $p_0$ can be related to $\Delta T$ through (*28*):

$$p_0 = \Gamma \rho C_V \Delta T \left(\frac{\tau_{relax}}{\tau_{pulse}}\right) \tag{3}$$

$$\tau_{relax} = \frac{r_{focus}}{v_a} \tag{4}$$

Where $\Gamma$ is Gruneisen parameter, $C_V$ is constant volume heat capacity, $\tau_{relax}$ is the acoustic relaxation time, $\tau_{pulse}$ is the pulse duration, $r_{focus}$ is the radius of probe focus. With 1725 nm excitation wavelength and 1.0 objective numerical aperture (NA), $r_{focus} = 526\ nm$ and $\tau_{relax} = 0.35\ ns$. Because the acoustic relaxation time is shorter than the pulse duration ($\tau_{pulse} = 10\ ns$) while the thermal relaxation time is longer, the heat can accumulate during the excitation while the pressure keeps propagating out. This difference leads to a distinct contribution of PT and PA in the optically detected signal. Assuming olive oil as the sample, we have $\alpha = 0.00043\ K^{-1}, v_a = 1490\ m/s, \eta \approx 0.3, n_0 = 1.42, \Gamma = 0.9, C_V = 1970\ J * K * kg^{-1}$. With Equations 1 to 5, the intensity ratio between the optically probed PT and PA signal can be calculated as:

$$\frac{\delta n_{PT}}{\delta n_{PA}} = \frac{2\alpha v_a^3 \tau_{pulse}}{\eta n_0^3 \Gamma C_V r_{focus}} \approx 37 \tag{5}$$

The theoretical analysis indicates the PT contribution is more than one order larger than the PA contribution. To validate this result, we measured PA and PT signals simultaneously using olive oil as a standard sample.

**The optically detected photothermal signal is 63 times larger than the photoacoustic signal.**

Using the SWIP microscope shown in **Fig. 1**, we performed fast digitization of single-pixel signal from olive oil (**Fig. 2**). The PT and PA contributions can be differentiated by their temporal profiles. Because the calculated acoustic relaxation time is shorter than the pulse duration, the PA initial pressure rise should have the same duration as the excitation pulse, which is 10 ns. In contrast, the PT signal should have a long exponential decay as indicated by the Newton's law of cooling. When the two focuses were tight and in a good lateral overlapping (**Fig. 2a**), the PT signal was found to overwhelm the PA signal (**Fig. 2b**). When zooming in the bounding box area of **Fig. 2b**, a bipolar PA oscillation was observed but with an amplitude 63 times smaller than that of PT (**Fig. 2c**).

In order to confirm that the initial oscillation signal was indeed from PA, we acquired 3 other signal traces under different focusing conditions. Because the acoustic wave propagates much faster than the heat, we can consider the PT as a locally confined signal. Therefore, enlarging probe focus size (**Fig. 2d**) or shifting probe focus laterally out of the excitation focus (**Fig. 2g** and **2j**) should selectively detect the PA signal. As expected, with the modified schemes, the relative amplitude of PA became larger in the detected signal (**Fig. 2e, 2h, 2k**), which could be clearly observed in the zoom-in views (**Fig. 2f, 2i, 2l**). In an extreme case, when the probe focus had zero overlap with the excitation focus, the PT contribution in the probed signal was eliminated and a typical acoustic bipolar oscillation was observed (**Fig. 2 j-l**). This result confirms the PA origin of the probed signal shown in **Fig. 2l**. The width of the strongest oscillation band in **Fig. 2f**, **2i**, **2l** was measured to be around 10 ns, consistent with the initial pressure rise theory.

We noticed a 2 times mismatch between the theoretical calculation and the experimental result. We attribute this mismatch to the different energy transfer efficiency between heat and acoustic and the slightly insufficient bandwidth of our photodetector. This mismatch does not influence the conclusion that the PT signal has more than one order larger amplitude than the PA signal in our focusing configuration (**Fig. 2a**).

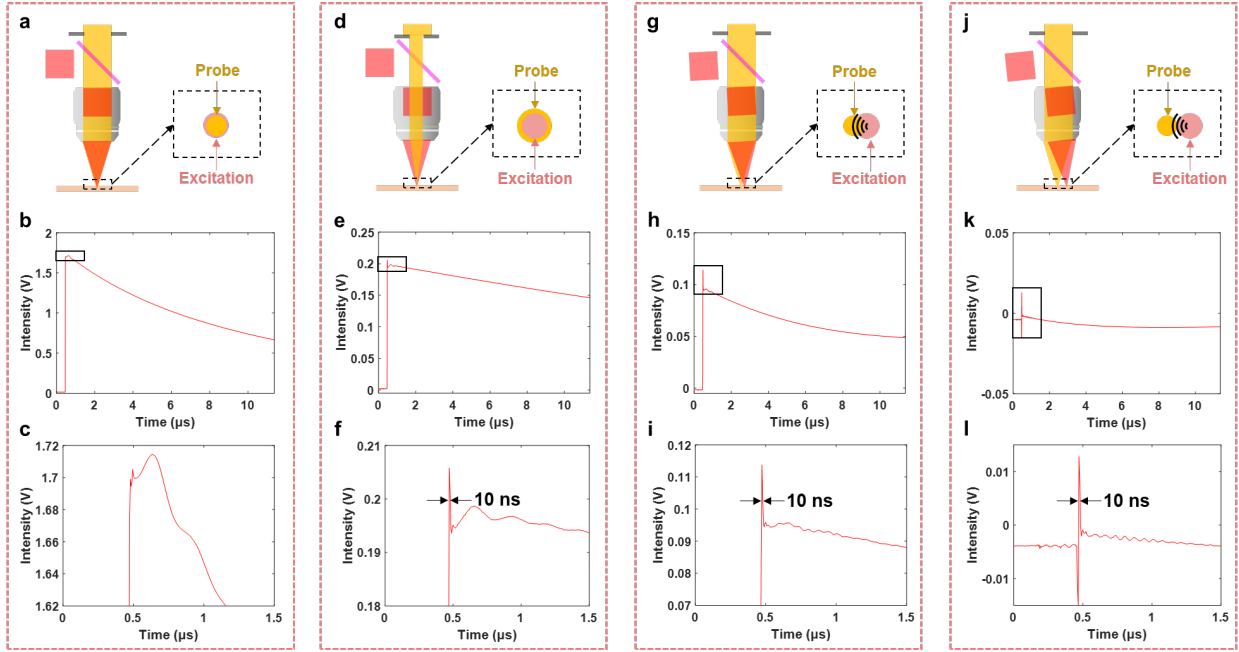

**Figure 2. Comparison of optically detected photothermal and photoacoustic signal.** (a) Normal SWIP focusing configuration when probe and excitation focuses have similar size and good lateral overlapping. (b) Signal trace under the configuration in (a). (c) Zoom-in of (b). (d) Focusing configuration where the probe focus is enlarged. (e) Signal trace of under the configuration in (d). (f) Zoom-in of (e). (g) Focusing configuration where probe focus has small lateral shift relative to the excitation focus. (h) Signal trace under the configuration in (g). (i) Zoom-in of (h). (j) Focusing configuration where probe focus has large lateral shift to the excitation focus. (k) Signal trace of under the configuration in (j). (l) Zoom-in of (k). Sample: Olive oil. SWIR excitation power on sample: 4.2 mW.

**A ball-lens model describing the optically probed photothermal signal**

To further interpret the dependence of SWIP intensity on the experiment parameters and focusing conditions, we have developed a ball-lens PT detection model (Supplementary section 1). When the size of the thermal lens matches with or is larger than the probe focus, the SWIP intensity calculated from the model is:

$$I_{SWIP} = A_1 \alpha \frac{1}{C_p} \eta \sigma N I_{probe} I_{excitation} \quad (6)$$

Here, $A_1$ is a constant; $\alpha$ is thermo-optic coefficient; $C_p$ is the heat capacity of the sample; $\eta$ is the percentage of the absorbed light power converted into heat; $\sigma$ is the overtone absorption cross section; $N$ is the number of molecules in the focus volume, which represents the molecular concentration given a fixed focus volume; $I_{probe}$ is the intensity of probe laser; and $I_{excitation}$ is the intensity of the excitation laser. Equation 6 shows a linear dependence of SWIP signal on laser intensity and molecular concentration, which is desirable for quantitative analysis.

**SWIP imaging sensitivity, spatial resolution, spectral fidelity, and linearity**

Knowing that PT is dominant in our detection, we optimized the detection scheme to fully utilize the slow-decaying PT signal (Describe in Materials and methods). Then, we characterized the system performance. **Fig. 3a** and **3b** show the XY section and YZ section of volumetric SWIP imaging of single 500 nm PS beads. The signal-to-noise ratio (SNR) of the XY image of bead was 25, which shows a high sensitivity of SWIP in detecting nano-objects. **Fig. 3c** and **3d** show the lateral and axial profile of a single 500 nm PS bead, where the lateral and axial FWHM were measured to be 0.92 µm and 3.5 µm. After deconvolution with the beads profile, the system's lateral and axial resolution were calculated to be 0.77 µm and 3.5 µm, which closely matched with the theoretical resolutions of 0.80 µm and 3.5 µm calculated with an objective NA of 1.0 and a wavelength of 1310 nm. Such a resolution is sufficient to resolve the subcellular features.

To confirm that the SWIP contrast is indeed from C-H overtone absorption, we acquired SWIP spectra of pure dimethyl sulfoxide (DMSO) and glycerol trioleate. The SWIP spectra shown in **Fig. 3e** and **3f** match the literature well (*11, 29*). **Fig. 3g** and **3h** show a linear dependence of the SWIP signal on both the excitation power and the molecular concentration, in consistence with the ball lens model. Collectively, **Fig.** 3e-h demonstrates the quantitative chemical analysis capability of SWIP microscopy.

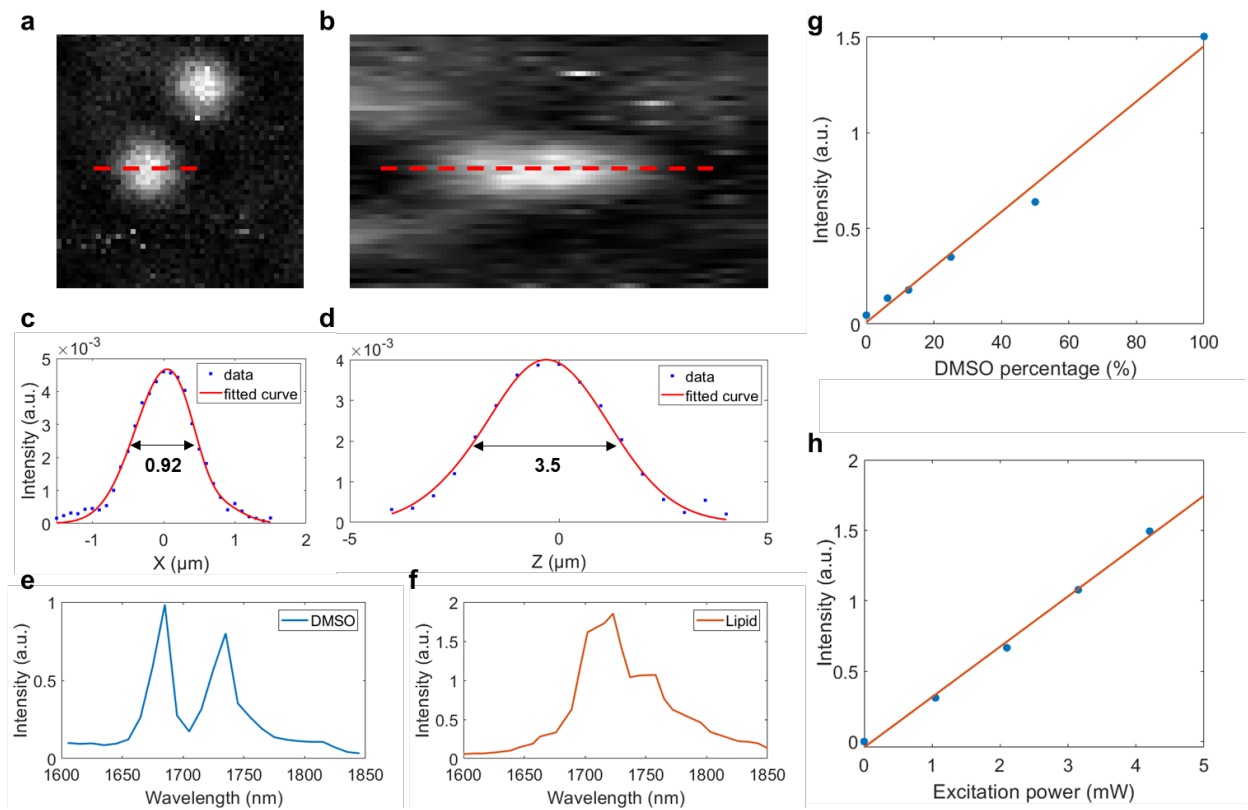

**Figure 3. SWIP microscope performance.** (a, b) XY and YZ section of volumetric SWIP image of single 500 nm PS beads. Excitation power on sample: 20 mW. (c, d) Single 500 nm PS bead's lateral and axial profile corresponding to dashed line in (a) (b). (e, f) SWIP spectrum of pure DMSO and glycerol trioleate. (g) SWIP signal dependence on concentration. Sample: DMSO solution in DMSO-D6. (h) SWIP signal dependence on excitation power. Sample: pure DMSO. Excitation power on sample: 4.2 mW.

## SWIP imaging of particles through a scattering phantom

The deep-penetrating capability of SWIP was first validated with scattering phantoms. Tissue-mimicking intralipid aqueous solution was placed between the objective and 1 μm PS beads dried on a coverslip (**Fig. 4a**). The thickness of the scattering medium was around 800 μm according to the objective working distance. **Fig. 4b** shows the SWIP imaging results through water, 1% intralipid, 5% intralipid or 10% intralipid. According to literature, 1% intralipid has a similar scattering coefficient to human skin epidermis (*30*). SWIP imaging of beads through 1% intralipid had a SNR of 259, which is close to the SNR of 297 in the pure water group. Through 5% intralipid, the image SNR dropped to 95 but still maintained a good quality. With 10% intralipid, the image SNR decreased to 6, but single 1 μm PS beads were still visible. As the scattering coefficient of 10% intralipid is 10 times as that of 1% intralipid (*31*), the imaging result through 10% intralipid indicated that SWIP has the capacity to reach millimeter-level penetration depth in biological tissue. To showcase the penetration advantage of SWIP, we performed near-infrared stimulated Raman scattering (SRS) imaging on the same samples. **Fig. 4c** presents the SRS imaging result with the same phantom and objective. In pure water condition, the SRS image showed higher resolution than the SWIP image due to shorter excitation wavelengths. However, the quality of SRS images quickly degraded as the intralipid concentration increased. No beads could be detected by SRS under 5% and 10% intralipid. Together, these data show that SWIP has the deep-penetrating capability through a highly scattering medium, which is beyond reach of a commonly used SRS microscope.

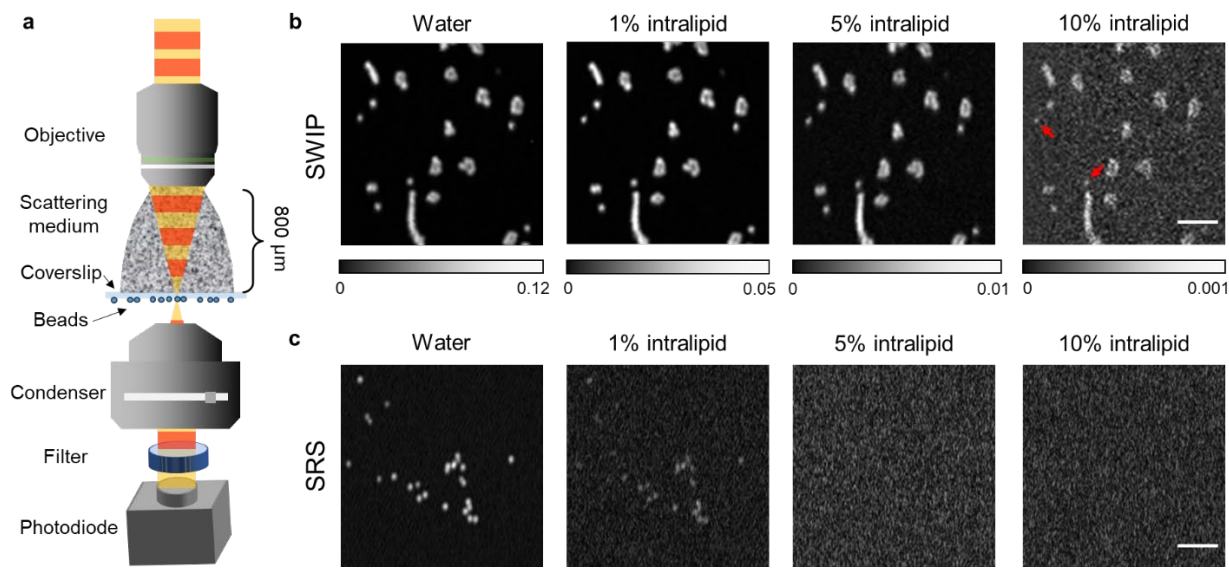

**Figure 4. SWIP imaging of PS particles through a highly scattering medium.** (a) SWIP schematic. (b) SWIP imaging results of 1 μm PS beads through water or scattering medium. Laser power on sample: 1725 nm: 20 mW. 1310 nm: 30 mW. (c) SRS imaging results of 1 μm PS beads through water or scattering medium. Laser power on sample: 798 nm: 10 mW. 1040 nm: 60 mW (water group); 798 nm: 50 mW. 1040 nm: 100 mW (intralipid groups). Scale bar: 10 μm. PS: polystyrene.

**Time domain extraction of small SWIP signal from water background**

After we obtained strong signals from polymer beads through a scattering medium, we demonstrated SWIP imaging of living cells in culture medium. As shown in **Fig. 5a**, the cell structure was correctly revealed by SWIP. Meanwhile, a background from water was also observed, which compromised the contrast of small intracellular lipids (red arrow). To address this issue, we took advantage of the different thermal decay characteristics between lipids and the bulk water. Because the SWIP heating volume is larger than the intracellular lipids especially in the axial direction, the heat dissipation of the small lipids is significantly faster than that of bulk water. The single-pixel SWIP trace in the lipid area (**Fig. 5b**) and water background area (**Fig. 5c**) supported this hypothesis. By fitting the SWIP signal with 2-component exponential decay, a significant difference in the thermal decay constant between lipid and background can be revealed. Two-component exponential fitting was selected considering that the SWIP signal originate from both in-focus target and out-of-focus background. By removing the first exponential component and applying soft thresholding according to the second decay coefficient (Supplementary Section 2), the water background was successfully removed, and the small lipids showed up more clearly (**Fig. 5d**). This method builds the foundation for imaging the cellular lipids in spheroid and tissue environments.

**SWIP imaging of intracellular lipids in an intact tumor spheroid**

Tumor-derived spheroid is a kind of in-vitro cancer model that better recapitulates tumor physiology and response than conventional two-dimensional culture (*32*). As cancer development is closely related to the altered lipid metabolism (*33*), imaging intracellular lipids inside spheroid can help understand cancer progression and test drug effectiveness. However, imaging cellular components inside a spheroid is challenging as the densely packed cells strongly scatter light. Sectioning (*34*) and tissue clearing (*35, 36*) have been applied to circumvent the strong scattering. Yet, the sectioning and clearing methods may alter the metabolic state of the spheroid and cannot be used for live sample study. Deep-penetrating multi-photon fluorescence microscopy and light-sheet fluorescence microscopy can image live spheroids (*37, 38*). However, the fluorescent labeling is perturbative, especially for small lipid molecules (*39*).

SWIP overcomes the above-mentioned challenges by label-free imaging of intracellular lipids in an intact spheroid. **Fig. 5e** shows SWIP imaging result of an intact tumor spheroid at 3 representative depths. In the raw SWIP images, the intracellular lipids were successfully identified at all three depths. After removing the water background, the intracellular lipids showed up cleanly. The cell nuclei can be identified as a negative contrast surrounded by intracellular lipids (blue arrow in **Fig. 5e**). Some hollow structure in the center of the spheroid can also be observed (white arrow in **Fig. 5e**). **Fig. 5f** is a background-removed three-dimensional rendering for the volumetric SWIP images of the spheroid. The volumetric image shows an enriched accumulation and a relative uniform distribution of lipid across the spheroid.

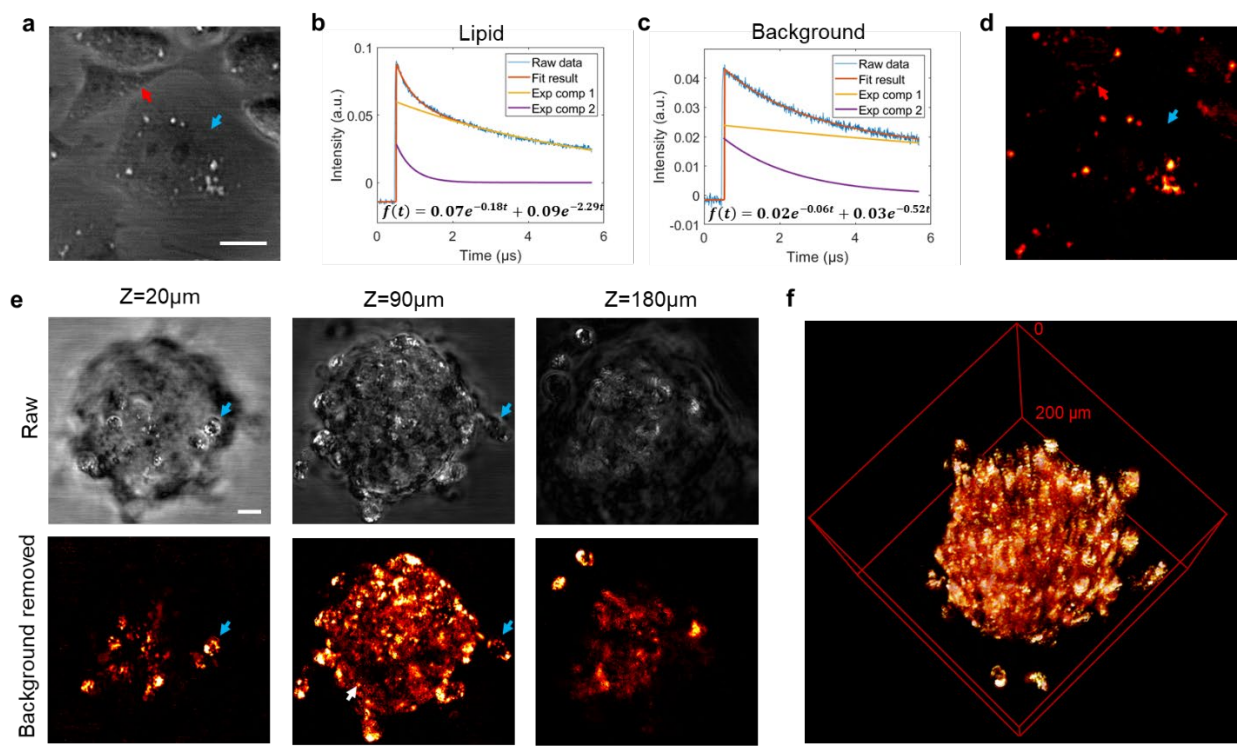

**Figure 5. SWIP imaging of cancer cells and spheroid.** (a) Raw SWIP image of monolayer OVCAR-5-cisR cells. (b) Single-pixel SWIP signal trace at the red arrow-pointed lipid area. (c) Single-pixel SWIP signal trace at the blue arrow-pointed background area. (d) Background rejection result of (a) using the decay characteristic. (e) SWIP imaging of an OVCAR-5-cisR spheroid. (f) 3-D rendering of volumetric SWIP imaging of OVCAR-5-cisR spheroid after background rejection. Laser power on sample: 1725 nm: 20 mW, 1310 nm: 30 mW. Scale bar: 20 µm.

**SWIP imaging of lipids in biological tissues**

Lipids play an important role in biological tissue including energy storage, signaling, protection function and transport of fat-soluble nutrients (*40*). Imaging lipid content and its distribution inside a tissue thus can enable a range of applications (*41*). Because fluorescence labeling is perturbative for the lipid molecules, vibrational imaging is widely adopted for lipid studies (*39*). As reviewed in the introduction, current vibrational imaging modalities do not allow high-resolution lipid imaging in deep tissue. Consequently, tissue sectioning is generally applied to allow high-resolution layer-by-layer imaging, but the sectioning process usually introduces morphological artifacts and often causes lipid loss (*42*).

To overcome the above-mentioned challenges. We explored SWIP imaging of lipids in various types of tissues. **Fig. 6a** shows the SWIP images of a fresh swine liver slice. The lipid and liver morphology revealed by SWIP was consistent with previously reported SRS results (*43*). Lipid droplets as small as 1 µm in diameter can be distinguished even at 300 µm deep inside the fresh liver with high blood content, which cannot be achieved via existing modalities. **Fig. 6b** shows the SWIP images of a mouse ear. Characteristic layered structures were observed, including hair at Z=0 µm, sebaceous gland at Z = 52 µm, and subcutaneous fat layer or cartilage at Z=156 µm. **Fig. 6c** reports the imaging result on a mouse brain slice with a thickness of around 1 mm. SWIP can image through the whole brain slice and well capture the myelin fibrous structure. **Fig. 6d** demonstrates SWIP imaging on a breast biopsy of a healthy human. Different layers of fat cells can be resolved across a 600 µm thick breast tissue.

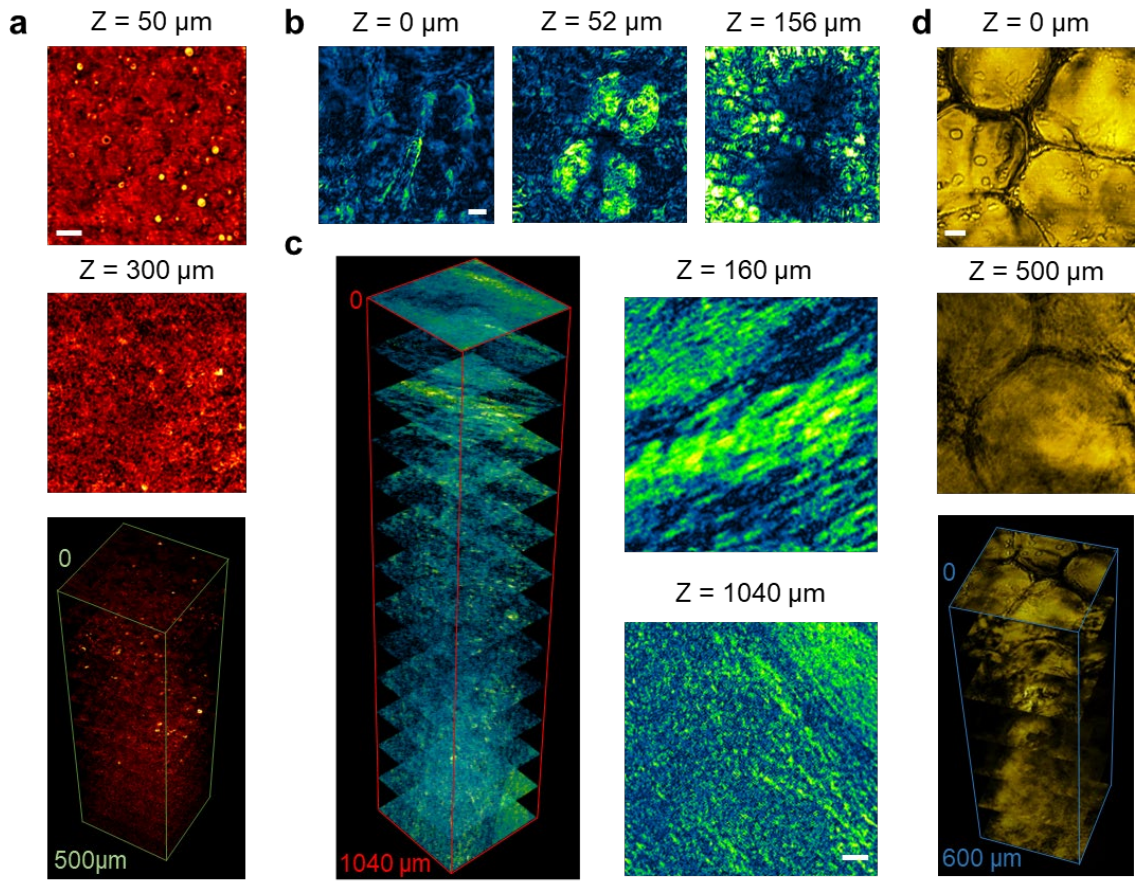

**Figure 6. SWIP imaging of lipids in biological tissues.** (a) SWIP imaging of fresh swine liver slice. (b) SWIP imaging of a mouse ear. (c) SWIP imaging of a mouse brain slice. (d) SWIP imaging of a breast biopsy from a healthy human. Laser power on sample: 1725 nm: 20 mW, 1310 nm: 50 mW. Scale bar: 20 μm.

Finally, towards in vivo deep-tissue imaging, we have built and tested an epi-detected SWIP system shown in **Fig. 7**. A major challenge facing epi-detection is the weak back-scattered probe laser intensity, which gives a poor SNR when a normal biased photodiode is used. To address this issue, we harnessed an amplified photodiode of higher sensitivity and achieved high-performance epi-detection. **Fig. 7b** shows an epi-detected SWIP image of 1-μm PS beads placed under 1% intralipid. The epi-SWIP is able to map PS the beads under an 800 μm-thick scattering phantom with a SNR of 19. **Fig. 7c** shows an epi-detected SWIP image of the sebaceous gland of mouse ear at the depth of 60 μm. The epi-imaged sebaceous gland shows a good match with the forward result (**Fig. 6b**). **Fig. 7d** reports epi-detected SWIP images of an intact mouse brain. Cell-like features were observed at the depth of 300 μm. The bright myelin-like features appeared at the depth of 500 μm, where the imaging plane passes from the cell-rich grey matter to the lipid-rich white matter. Together, these results show that in-vivo SWIP imaging could be potentially achieved.

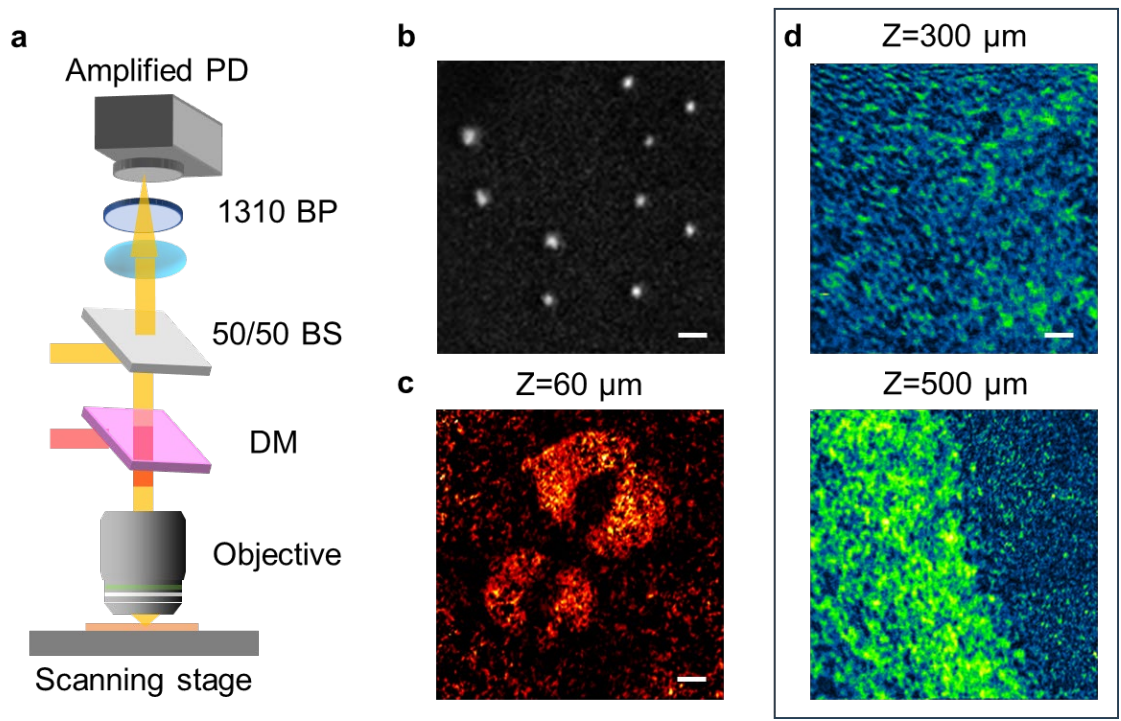

**Figure 7. Epi-detected SWIP imaging.** (a) Epi-detection schematic. PD: Photodiode, BP: Bandpass filter, BS: beam splitter, DM: dichroic mirror (b) Epi-SWIP imaging of 1 μm PS beads under 800 μm-thick 1% intralipid. Scale bar: 5 μm. (c) Epi-SWIP imaging of mouse ear at 60 μm depth. Scale bar: 20 μm. (d) Epi-SWIP imaging of mouse brain at 300 μm and 500 μm depth. Scale bar: 20 μm. Power on sample: 1725 nm: 20 mW. 1310 nm: 53 mW.

**DISCUSSION**

We have developed SWIP microscopy to bridge the gap for deep-tissue vibrational imaging at subcellular resolution. We showed the PT amplitude was 63 times stronger than PA in the optically detected signal when the two SWIR focuses are on similar size and good overlapping. The strong PT signal enabled a high detection sensitivity to resolve single nanoparticles underneath a strongly scattering medium. Remarkably, SWIP achieved millimeter-level vibrational imaging depth through highly scattering intralipid phantom, mouse brain slice, and human breast biopsy. SWIP further enabled intracellular imaging across the whole tumor spheroid and lipid mapping in various tissue including liver, skin, brain, and human breast.

A comparison of SWIP with existing vibrational imaging modalities in the dimension of penetration depth versus spatial resolution is illustrated in **Fig. 8**. These techniques can be clustered into three groups. The first group includes coherent Raman scattering (CRS), mid-infrared photothermal (MIP), confocal Raman microscopy (CRM) in the bottom-left. These modalities have subcellular resolution and sensitivity but limited imaging depth on the scale of tens of microns to 100 microns. The second group includes shortwave infrared photoacoustic microscopy (SWIR PAM), shortwave infrared diffuse optical imaging (SWIR DOI) and spatial offset Raman spectroscopy (SORS) in the top-right. These modalities have deep penetration depth but relatively poor spatial resolution on the scale of a hundred of microns to several millimeters. Although SWIR PAM can achieve higher spatial resolution with a tight optical focus, the large signal loss eventually prevents SWIR PAM to probe small intracellular components. Clearly, there exists a gap between the first group and the second group for deep tissue vibrational imaging at subcellular resolution. As demonstrated in this work, SWIP successfully fills in the gap between the first and second groups. The millimeter-deep, micron-resolution, high-sensitivity vibrational imaging capability provided by SWIP opens exciting opportunities for many applications including spheroids study, slice-free tissue pathology, embryo imaging, etc.

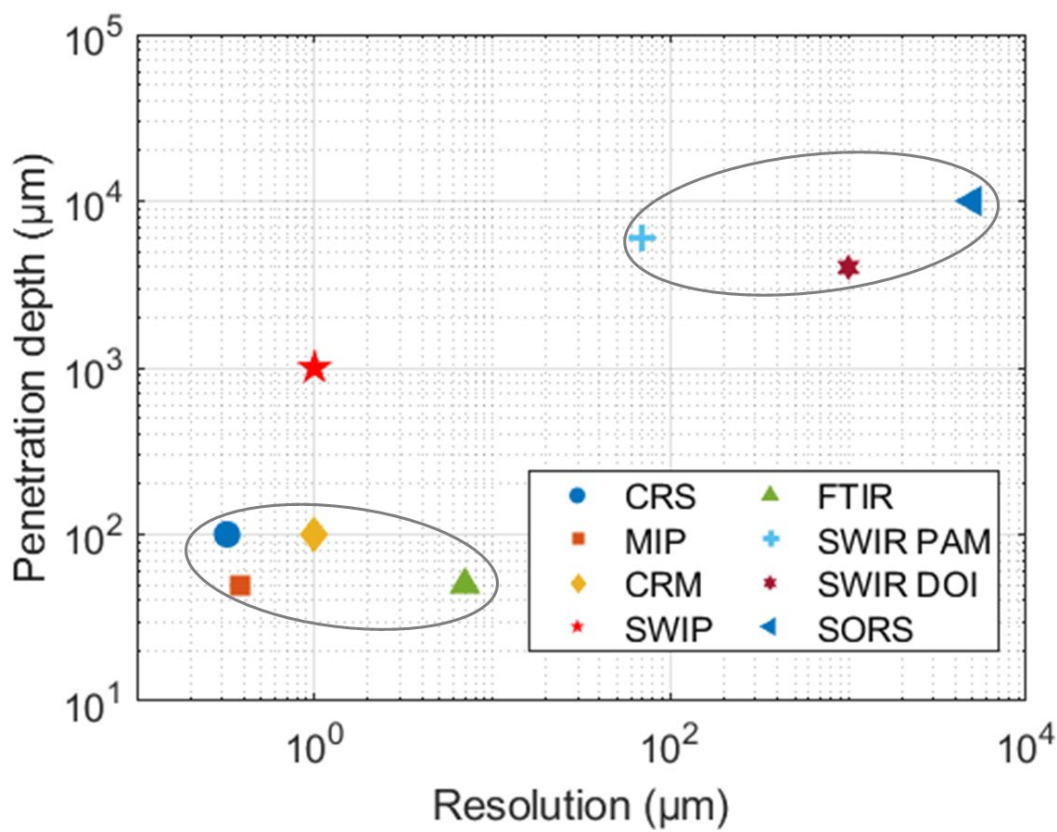

**Figure 8. Penetration depth versus spatial resolution of vibrational imaging modalities**. The data of penetration depth and spatial resolution are collected from reference (*3, 4, 6, 15, 16, 44*). CRS: coherent Raman scattering, MIP: mid-infrared photothermal, CRM: confocal Raman microscopy, FTIR: Fourier transform infrared spectroscopy, SWIR PAM: short-wave infrared photoacoustic microscopy, SWIR DOI: shortwave infrared diffusive optical imaging, SORS: spatial-offset Raman spectroscopy, SWIP: shortwave infrared photothermal microscopy.

It is noteworthy that the SWIP signal is sensitive to the focusing condition. As shown in **Fig. 2**, the best configuration for PT is when the two focuses have similar size and in good XY overlapping. The PA signal is favored when the probe volume is larger than the excitation volume or when the two focuses have large offsets, which is the case in reported photoacoustic remote sensing (*18*). Compared to PT, PA has its own merits. The PA signal is closely related to the mechanical property which is valuable for many applications (*45*). Furthermore, the high-frequency characteristics of PA can circumvent low-frequency noises, which can give a higher

SNR in certain cases. Our data and theory relating the signal level with the focusing condition can guide future system design for better utilization of the different strengths of PT and PA processes.

The reported SWIP microscope can be improved in several aspects. By far, our single-color SWIP imaging using the first overtone vibration at 1725 nm mainly targets the C-H bond that is enriched in lipid content. With a wavelength-tunable excitation laser, hyperspectral SWIP imaging and subsequent decomposition can differentiate multiple molecular species such as proteins, lipids, cholesterol with chemical specificity (*11, 14*). Second, the acquisition time of current SWIP microscope is around 3 minutes per frame, limited by the 2 kHz excitation laser repetition rate. Combining a higher-repetition-rate excitation laser with galvo scanning, the imaging speed of SWIP can be much improved for in-vivo applications. Third, SWIP has better performance imaging through homogeneous scattering phantom but encounters image degradation at hundreds-of-micrometers deep in the tissue. We attribute such degradation to tissue-induced aberration which inevitably distorts the laser focus. By implementing adaptive optics for aberration correction (*46*), SWIP can reach even deeper into the tissue. Lastly, we envision that SWIP microscopy can be upgraded into SWIP-OCT for high-speed volumetric vibrational imaging of organoids and tissues.

# MATERIALS AND METHODS

## Chemicals

The oil sample was prepared by sandwiching 2 μl vegetable oil with two no.1 coverslips. An 80 μm double sided tape was placed between the two coverslips as a spacer. To prepare the polystyrene (PS) beads sample, an aqueous PS beads solution was first prepared and mixed well with an ultrasound homogenizer to avoid large aggregates. Then the beads solution was dried on a no.1 coverslip. When performing imaging, the coverslip side was on the top to isolate the beads from the immersion medium or the scattering phantom.

## Cancer cells and spheroids

Cisplatin-resistant ovarian cancer cells (OVCAR-5-cisR) (*47*), were prepared by the Daniela Matei Lab at Northwestern University. The cells were cultured in RPMI 1640 with 10 % fetal bovine serum, 100 units/ml penicillin and 100 μg/ml streptomycin. Cells were seeded on a coverslip for 24 hours for monolayer cell imaging. To form ovarian cancer spheroids, 200 μl/well of OVCAR-5-cisR cell suspension were added to an ultra-low attachment 96 well plate with a cell density of $0.5 \times 10^4$ cells/ml. The spheroids were cultured for 7 days. Cells were then kept in 1x Phosphate Buffered Saline (PBS) and sealed in between two coverslips with spacers.

## Biological tissues

The fresh swine liver was purchased from the supermarket. Before imaging, the liver was hand-sliced into around 3 mm and sandwiched between two no.1 coverslips. The mouse ear sample was isolated from a 6-month-old mouse and fixed with 10% formalin solution. Before imaging, the ear was rinsed with 1x PBS and then attached to a coverslip. The mouse brain was from a 6-month-old mouse and fixed with 10% formalin solution. The brain was sliced to be 1-mm thick with a vibratome. Before imaging, the brain was rinsed with 1x PBS then attached to a coverslip. The de-identified healthy human breast biopsy sample was obtained from Susan G. Komen tissue Bank at the IU Simon Cancer Center. The breast biopsy sample was freshly frozen and had a thickness of around ~ 0.5 mm. Before imaging, the breast biopsy sample was defrosted then sandwiched between two coverslips.

## SWIP microscope

Both pump and probe beams are in the shortwave infrared window to ensure deep tissue penetration. The pulsed pump beam is generated by an optical parametric oscillator (DX-1725-OPO, Photonics Industries), with a repetition rate of 2 kHz, wavelength centered at 1725 nm, and a pulse duration of 10 ns. The 1310 nm probe beam is provided by a CW diode laser (TURN-KEY CCS-LN/1310LD-4-0-0/OC, Research Lab Source Corporation). The beam sizes of the two lasers were adjusted to be around 6 mm with lens pairs. After beam expansion, the two lasers were collinearly combined with a dichroic mirror and then focused into a sample through an objective lens. Two objectives were used for the experiments. A 1.0 NA water-immersion objective with 800 μm working distance was used to image the polymer beads, cancer cells, spheroid, swine liver, and human breast biopsy. A 2 mm working distance objective with an effective NA of ~0.4 was applied for mouse ear and brain imaging. The transmitted light from the sample was collected by an air condenser (D-CUO, Nikon) with an adjustable aperture. After the condenser, the remaining excitation laser was filtered out by a 1310 nm bandpass filter (FBH1310-12, Thorlabs). The signal-carrying probe beam is detected by a biased InGaAs photodiode (PD). When recording both PA and PT signals, we used a small-area high-speed PD (70 MHz bandwidth, 0.8 $mm^2$, DET10N2, Thorlabs). When only targeting the PT signal, we used a slower PD with larger active area (11.7 MHz bandwidth, 3.14 $mm^2$, DET20C2, Thorlabs). The photocurrent from the PD is converted to a voltage signal with a 50 Ohm impedance and then amplified by an AC-coupled low-noise voltage amplifier (100 MHz bandwidth, SA230F5, NF corporation) and digitized by a high-speed data acquisition card at 180 MSa/s (ATS9462, Alazar Tech). Every excitation laser pulse corresponds to a single pixel in the image. The image was formed by sample scanning achieved with a stage (Nano-Bio 2200, Mad City Labs). The volumetric image was acquired with a motorized z-knob to allow axial scanning.

**SWIP image formation**

Every pixel in the SWIP image corresponds to one excitation laser pulse, for which, a temporal trace of the probe laser intensity will be recorded. A gating method is applied to turn the signal temporal trace to pixel intensity. Two averaging windows are used: the first window is set before the excitation pulse arriving to estimate the probe intensity baseline; the second window starts at intensity extremum right after the excitation to obtain the changed probe intensity. The pixel intensity is assigned to be the difference between the two window averages. This gating method takes advantage of the long decay of PT signal for a better SNR. Multiple window sizes have been tested and the size of 80 sampling points (~450 ns) is chosen to output the best SNR.

**SWIP spectroscopy**

The SWIP spectra were acquired by replacing the single-color excitation laser with a tunable excitation laser (Opolette HE 355 LD, OPOTEK Inc), which has a tuning range from 410 to 2400 nm, a pulse duration of 5 ns and a repetition rate of 20 Hz. Other parts in the SWIP microscope remained unchanged. The spectral scanning was achieved by manually tuning the laser wavelength with a step size of 10 nm.

**SRS microscope**

Two synchronized femtosecond laser pulse trains with an 80 MHz repetition rate were used for SRS imaging. The wavelengths of the lasers are at 800 nm and 1040 nm to target the C-H stretch vibration. The 1040 nm laser is modulated by an acousto-optic modulator (AOM) at 2.27 MHz to separate the SRS signal from the laser repetition rate frequency. The SRS is conveyed by the modulation transfer from 1040 nm to 800 nm laser at 2.27 MHz. The two lasers are chirped with SF57 glass rods for spectral focusing which provides a high spectral resolution. A dichroic mirror spatially combines the two lasers. The combined beam passes a pair of galvo mirrors for laser scanning, then is focused on the sample by the same objective (1.0 NA, 800 μm working distance) used for SWIP. The transmitted light is collected by a 1.4 NA oil-immersion condenser and filtered by a 980 nm short pass filter. The residue 800 nm laser is detected by a biased photodiode. The SRS signal is obtained by demodulating the signal from a photodiode with a lock-in amplifier.

**ACKNOWLEDGEMENTS:** Human breast samples from the Susan G. Komen Tissue Bank at the IU Simon Cancer Center were used in this study. We thank Dr. Lu Lan and Jiaze Yin for the helpful discussion on the photothermal detection scheme.

**Funding:** This work is supported by NIH R35GM136223, R33CA261726, and R01 HL125385 to JXC.

**Author contributions:** HN developed the SWIP system, designed experiments, processed data, built theoretical models, and drafted the manuscript. YY helped carry out the experiments, and manuscript writing, performed the theoretical analysis and prepared cancer cell and spheroid samples. ML provided the mouse tissue and helped in data analysis. YZ contributed to the project formulation and building of theoretical models. XG contributed to the project formulation and data analysis. CP prepared the human breast biopsy sample. LW helped in the SRS experiment. JXC initialized the project and provided scientific guidance.

**Competing interests:** The authors declare that they have no competing interests.

**Data availability:** All data necessary to evaluate this work is shown in the paper and the Supplementary Materials.

# Supplementary Materials

## Millimeter-deep micron-resolution vibrational imaging by shortwave infrared photothermal microscopy

Hongli Ni, Yuhao Yuan, Mingsheng Li, Yifan Zhu, Xiaowei Ge, Chinmayee Prabhu Dessai, Le Wang, and Ji-Xin Cheng[*]

*Corresponding author. Email: jxcheng@bu.edu

**This PDF file includes:**

    Supplementary Text
    Figs. S1 to S2
    Captions of Movies S1 and S2
    References (48 to 49)

**Other Supplementary Materials for this manuscript include the following:**
    Movies S1 to S2

# Supplementary Text
## 1. A ball-lens model for the optically detection photothermal signal

To better interpret the shortwave infrared photothermal (SWIP) contrast, we developed a straightforward ball-lens model which connects the SWIP intensity with the thermal-induced refractive index change (**Fig. S1**). Our model is based on a geometrical approximation, which ignores the interference and diffraction for ease of understanding. A more rigorous analysis can also be derived based on scattering theories (*48, 49*).

Before introducing the ball-lens model, we first calculated the thermal-induced refractive index change:

$$\delta n = \alpha \Delta T = \alpha \frac{1}{C_p} \eta \sigma N \tau I_{excitation} \qquad (1)$$

Where $\alpha$ is thermo-optic coefficient, $\Delta T$ is the temperature change induced by the photon absorption; $C_p$ is the heat capacity of the sample; $\eta$ is the percentage of absorbed light power converted into heat; $\sigma$ is the overtone absorption cross section; $N$ is the number of molecules in the focus volume; which represents the molecular concentration given a fixed focus volume; $\tau$ is the duration of the excitation pulse. As the excitation pulse is short in SWIP, we do not need to consider the heat dissipation here; $I_{excitation}$ is the average intensity of the excitation laser within the pulse duration.

**Fig. S1a** shows the case when the thermal-induced lens is larger than or matches with the size of the probe laser. The generated thermal lens can be simplified as a ball lens with a uniform refractive index change $\delta n$ and a radius $r$. $n$ is the original refractive index of the sample, $p, q$ are the distances from the thermal lens to the unmodulated probe focus and the modulated probe focus, respectively. $\theta_1, \theta_2$ are the divergence angles of the unmodulated and modulated probe laser. $a$ is the radius of the detection aperture. $a_1, a_2$ are the radius of the unmodulated and modulated probe beam at the plane of detection aperture. According to the ball lens formula, we have:

$$\frac{1}{f} = \left(\frac{n - \delta n}{n} - 1\right)\frac{2}{r} = \frac{2\delta n}{nr} \qquad (2)$$

Assume the probe laser is a Gaussian beam. According to Gaussian beam thin-lens equation:

$$\frac{\theta_1}{\theta_2} = m = \frac{f}{\sqrt{(p-f)^2 + z_R^2}} \qquad (3)$$

Where $z_R$ is the Rayleigh range of the unmodulated probe laser. As $\delta n \ll n$, we have $|f| \gg r \sim z_R$. Therefore,

$$\frac{\theta_1}{\theta_2} \approx \frac{f}{f - p} \qquad (4)$$

Assume a constant total intensity $I_{probe}$ before and after the thermal modulation. Denote the intensity distribution on the detection aperture plan for the unmodulated and modulated probe beam as $I_1$ and $I_2$:

$$I_1(r) = I_{probe} \frac{2}{\pi a_1^2} \exp\left(\frac{-r^2}{a_1^2}\right) \qquad (5)$$

$$I_2(r) = I_{probe} \frac{2}{\pi a_2^2} \exp\left(\frac{-r^2}{a_2^2}\right) \qquad (6)$$

The transmitted laser intensity after the detection aperture can be then written as follows assuming $a \ll a_1, a_2$:

$$I_1^{det} = \int_0^a I_1(r) * 2\pi r\, dr = I_{probe}\left(1 - \exp\left(-\frac{a^2}{a_1^2}\right)\right) \approx I_{probe}\frac{a^2}{a_1^2} \tag{7}$$

$$I_2^{det} = \int_0^a I_2(r) * 2\pi r\, dr = I_{probe}\left(1 - \exp\left(-\frac{a^2}{a_2^2}\right)\right) \approx I_{probe}\frac{a^2}{a_2^2} \tag{8}$$

The detected modulation depth induced by the thermal lens can then be calculated as:

$$\left(\frac{\Delta I}{I}\right)_{e\approx b} = \frac{I_2^{det} - I_1^{det}}{I_1^{det}} = \frac{a_1^2}{a_2^2} - 1 \tag{9}$$

Under paraxial approximation, $a_1 \approx \theta_1 Z$, $a_2 \approx \theta_2 Z$. Also utilize $|f| \gg r \sim p$, Equation 9 can be re-written as:

$$\left(\frac{\Delta I}{I}\right)_{e\approx b} = \frac{a_1^2}{a_2^2} - 1 = \frac{\theta_1^2}{\theta_2^2} - 1 = \left(\frac{f}{f-p}\right)^2 - 1 = \frac{-2fp - p^2}{(f-p)^2} \approx -2\frac{p}{f} = -4\frac{\delta np}{nr} \tag{10}$$

Here we focused on the modulation depth because the probe intensity and the aperture size can be easily adjusted to change the absolute detected laser intensity. In our scheme, the modulation depth is the decisive factor for signal-to-noise ratio. The absolute SWIP intensity can be written as follows, where $\beta$ is a constant representing the probe laser transmission from the sample to the detector.

$$I_{SWIP} = \left(\frac{\Delta I}{I}\right)_{e\approx b}\beta I_{probe} = -4\frac{\delta np}{nr}\beta I_{probe} \tag{11}$$

Combining equation 1 and 11, we finally get:

$$I_{SWIP} = A_1 \alpha \frac{1}{C_p}\eta \sigma N \tau I_{probe} I_{excitation} \tag{12}$$

Where $A_1$ is a constant summarizing the contributions of $n$, $r$, $\beta$. Equation 12 indicates a linear relationship between SWIP intensity and molecular concentration as well as incident laser intensity, which is desirable for quantitative chemical analysis.

We also analyze the case when the thermal lens is smaller than the probe focus, which is the situation described in the main text **Fig. 2d**. As shown in **Fig. S1b**, the enlarge of the probe focus results that a smaller portion of light is modulated by the thermal lens. This equivalent to a reduced $I_{probe}$ for calculating $I_2^{det}$. Denotes the probe beam radius is $w$ at the plane of the thermal lens. The portion of the modulated probe laser and the new $I_2^{det}$ can be calculated as:

$$I_{probe}^{mod} = \int_0^r I_{probe}\frac{2}{\pi w^2}\exp\left(\frac{-2R^2}{w^2}\right) * 2\pi R\, dR \approx I_{probe}\frac{r^2}{w_1^2} \tag{13}$$

$$I_2^{det} = I_{probe}^{mod}\frac{2a^2}{a_2^2} = I_{probe}\frac{r^2}{w_1^2}\frac{a^2}{a_2^2} \tag{14}$$

The modulation depth for the **Fig. S1b** can then be written as:

$$\left(\frac{\Delta I}{I}\right)_{e<b} = \frac{I_2^{det} - I_1^{det}}{I_1^{det}} = \frac{r^2}{w^2}\frac{a_1^2}{a_2^2} - 1 = \frac{r^2}{w^2}\left(\left(\frac{\Delta I}{I}\right)_{e\approx b} + 1\right) - 1 \approx \frac{r^2}{w^2}\left(\frac{\Delta I}{I}\right)_{e\approx b} \tag{15}$$

Equation 15 shows a smaller modulation depth when the thermal lens is smaller than the probe focus. The reduce of modulated portion can also be utilized to explain the PT modulation depth decrease when the thermal lens has a lateral offset to the probe focus. Noted here this model is based on geometrical derivation and considers no diffraction. A more rigorous analysis should be carried out with scattering theories. Overall, our theoretical model well explains our experiment result in the main text **Fig. 2**.

## 2. SWIP signal fitting and background rejection

The procedure to reject the water background is shown in **Fig. S2**. Every signal trace is first fitted with a two-component exponential function. Two-component exponential function is selected considering the detected SWIP signal should consist of the in-focus target and out-of-focus water background contribution. Then a weight w will be assigned according to the decay coefficient d of the faster decay among the two fitted components. The weight function is a soft-thresholding function with a cut-off decay constant around 1.5 µs. The final pixel intensity at the background removed image is calculated as the product of the assigned weight w and the amplitude c of the fitted fast decay component.

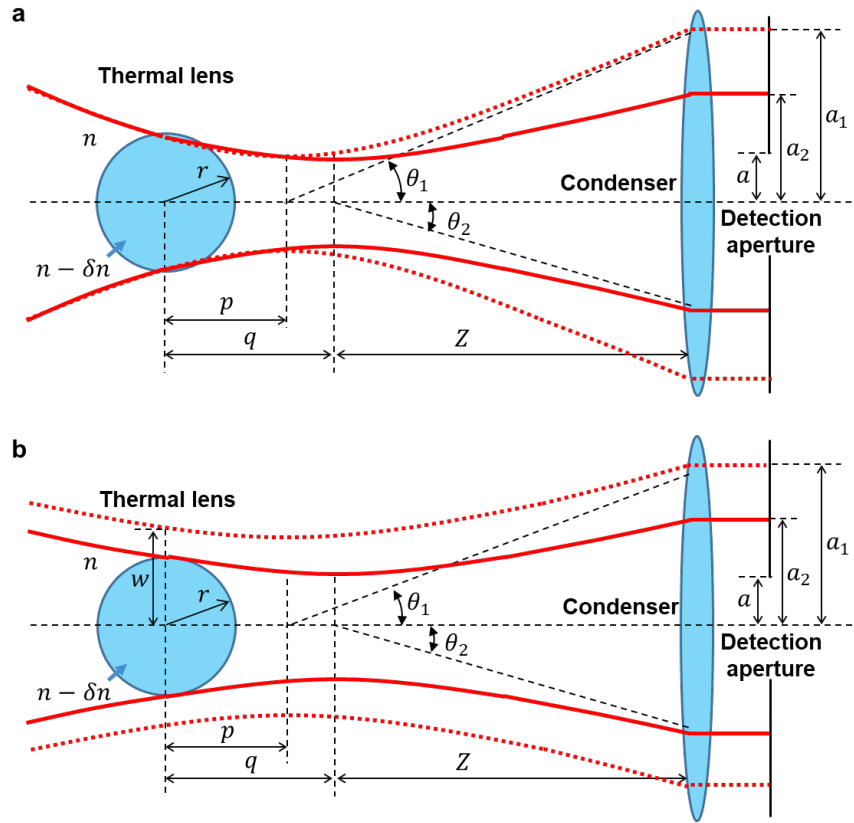

**Figure S1. A ball-lens model for the optically detected photothermal signal**. (a) Detection schematic when the thermal lens is larger or matches with the size of the probe focus. (b) Detection schematic when the thermal lens is smaller than the size of the probe focus.

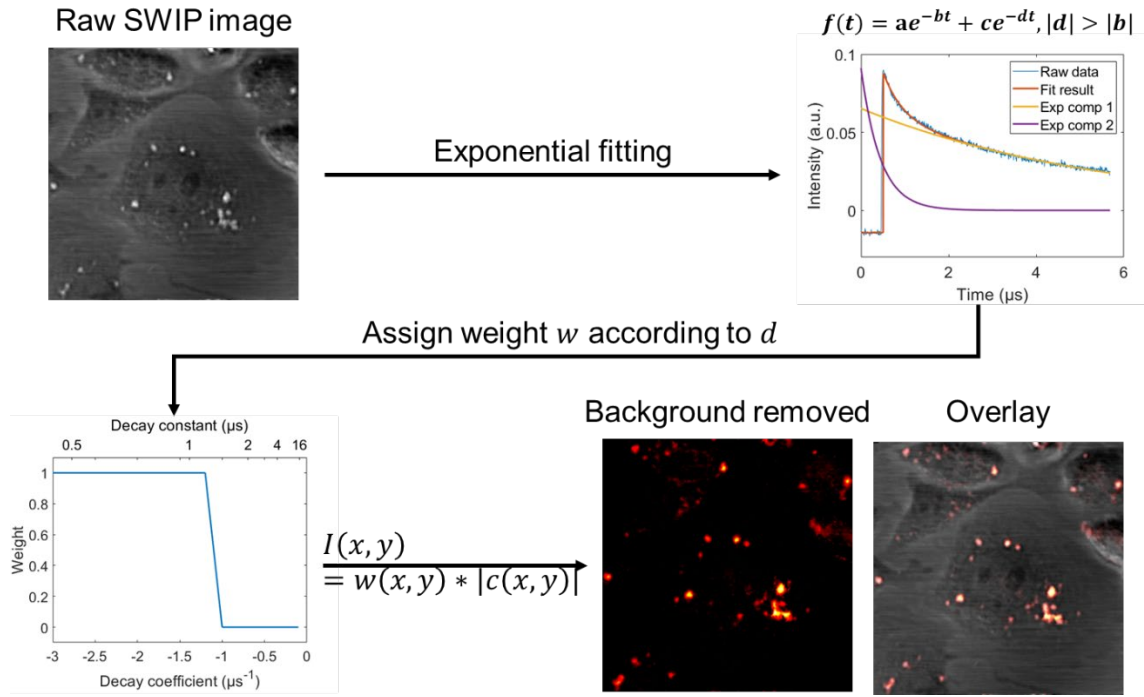

**Figure S2. Water background rejection workflow**.

**Movie S1. Layer-by-layer view of volumetric SWIP imaging of OVCAR-5-cisR spheroid.** Field of view: 200*200 μm.

**Movie S2. Three-dimensional rotation view of volumetric SWIP imaging of OVCAR-5-cisR spheroid.**